\documentstyle[aps,preprint]{revtex}
\begin{document}
\draft
\title{Charged particles in random magnetic fields and the critical
  behavior in the fractional quantum Hall effect}
\author{Bodo Huckestein}
\address{Institut f\"ur Theoretische Physik, Universit\"at zu K\"oln,
  Z\"ulpicher Stra\ss{}e 77, D-50937 K\"oln, Germany}
\date{\today}
\maketitle
\begin{abstract}
  As a model for the transitions between plateaus in the fractional
  Quantum Hall effect we study the critical behavior of
  non-interacting charged particles in a static random magnetic field
  with finite mean value. We argue that this model belongs to the same
  universality class as the integer Quantum Hall effect. The
  universality is proved for the limiting cases of the lowest Landau
  level, and slowly fluctuating magnetic fields in arbitrary Landau
  levels. The conjecture that the universality holds in general is
  based on the study of the statistical properties of the
  corresponding random matrix model.
\end{abstract}

The integer (IQHE) and fractional quantum Hall effects (FQHE) show
remarkable similarities despite the differences in their origin. While
the fundamental excitation gap is due to the strong magnetic field in
the IQHE \cite{Lau81}, strong Coulomb correlations are responsible for
the gap in the FQHE \cite{Lau83}. However, in both effects the
localization of electrons and quasiparticles, respectively, is
believed to be responsible for the formation of the plateaus in the
Hall conductivity \cite{AA81,Hal82,Lau83}. At the transitions between
successive plateaus in the IQHE scaling behavior has been observed
\cite{WTPP88,Huc95r}. This has been successfully interpreted as a
disorder-induced localization-delocalization transition for
non-interacting electrons \cite{Pru88,Huc95r}. Most remarkably,
it was found experimentally that the temperature dependent scaling
behavior of the transition between the FQH plateaus at filling factors 1/3
and 2/5 is described by the same scaling exponent as the transitions
between integer quantum Hall plateaus \cite{EWTS90}. Similar agreement
was obtained for the transition from filling factor 2/3 to 1
\cite{KHKP91}.

A theoretical description that makes the similarity between integer
and fractional QHE explicit is the ``composite fermion'' (CF) theory
of the FQHE \cite{Jai89}. It relates states of the interacting
electron system at filling factor
\begin{equation}
  \label{law}
  \nu=\nu'/(\nu'p\pm1)
\end{equation}
to states of non-interacting electron at filling factor $\nu'$ by
attaching an even number $p$ flux quanta to each electron. The magical
filling factors of the interacting electron system are interpreted as
filled Landau levels $(\nu'=\mathrm{integer})$ of the CFs. Based on
this approach, Jain, Kivelson, and Trivedi argued that transitions
between two FQH plateaus fall into the universality class of the IQHE
if these correspond to successive filled Landau levels of the CFs
\cite{JKT90}. The transitions for which scaling behavior was observed
correspond to the transitions from $\nu'=1$ to $\nu'=2$ for $p=2$ and
both signs in eq.~(\ref{law}) \cite{Jai89}.

Formally, the attachment of flux quanta can be achieved by the
introduction of a Chern-Simons vector potential \cite{LF93,HLR93}. In
a mean-field approximation this theory describes non-interacting CFs
in a uniform magnetic field corresponding to the filling factor $\nu'$
of the CFs. While on the mean-field level the universality of integer
and fractional QH transitions is thus manifest, the Chern-Simons field
is a dynamical gauge field and one has to worry about
the effects of its fluctuations. While not much is known about the
influence of the dynamics of the gauge field on the localization
properties, the effects of static fluctuations in the magnetic field
have recently attracted a lot of attention, in particular in the
context of the FQH system at filling factor 1/2. Static fluctuations
in the Chern-Simons field are due to static fluctuations in the
electron density that are induced by a residual disorder
potential. Most discussions in the literature focused on
non-interacting charged particles in a fluctuating magnetic field with
vanishing mean value relevant to the filling factor 1/2.
The results are rather controversial. Some authors
\cite{KZ92,KWAZ93,AHK93,KO95} claim to present evidence for a
localization-delocalization transition in contrast to the scaling
theory \cite{AALR79} according to which all states in two-dimensional
systems are localized in the absence of a strong magnetic field.
However, other authors, while observing a strong enhancement of the
localization length, find no true transition \cite{SN93,LC94,AMW94}.
We will consider the situation relevant to the transitions in the
FQHE. Since the FQH plateaus only form if the CF Landau levels are
well separated we will assume that the average magnetic field is
strong compared to the fluctuations. In this limit we will show that
the critical behavior is the same as that in the IQHE.

In this paper we treat the model of non-interacting charged particles
in a random magnetic field with a non-zero average magnetic field. If
the average magnetic field is strong compared to the fluctuations this
model describes the transitions between FQH plateaus if the charged particles
are thought off as non-interacting composite fermions. We will assume
that the average magnetic field is strong enough to neglect the
coupling between different Landau levels. Then there are two limits in
which the fluctuating magnetic field is strictly equivalent to a
random electrostatic potential. First, if only states of the lowest
Landau level are occupied for arbitrary correlations length of the
fluctuating magnetic field. Secondly, if the correlation length of the
fluctuations is large compared to the average cyclotron radius for
arbitrary Landau level index. The latter situation corresponds to the
semi-classical limit studied previously \cite{LCK94}. In general, a
fluctuating magnetic field is not equivalent to an electrostatic
potential. However, in the limit of well-separated Landau levels the
statistics of matrix elements of the random magnetic field Hamiltonian
and a random electrostatic potential are quite similar. While the
differences will be reflected in non-universal quantities like the
density of states, we conjecture that they do not lead to different
critical properties. This conjecture stands on the same footing
as the universality in the IQHE that has only been demonstrated
numerically for short correlation lengths in the two lowest Landau
levels and in the semi-classical limit of large correlation length
\cite{Huc95r}.

Our conclusions are based on the properties of the random matrix
model generated by projecting the Hamiltonian onto the Landau levels
of the average magnetic field $B_0$. The Hamiltonian $H$ containing
the Chern-Simons vector potential ${\bf a}$ can be expressed as a sum of the
Hamiltonian $H_0$ of the system with constant magnetic field
$B_0{\bf e}_z=\nabla\times {\bf A}$ and a part $H'$ due to the fluctuating
Chern-Simons field,
\begin{eqnarray}
  \label{Hamiltonian}
    H &=& \frac{1}{2m^*} \left({\bf p} -e{\bf A} - e {\bf a}\right)^2,\\
    H &=& H_0 + H',\\
    H_0 &=& \frac{1}{2m^*}\left({\bf p}-e{\bf A}\right)^2,\\
    H' &=& \frac{1}{2m^*} \left(-e \left(({\bf p}-e{\bf A}){\bf a}
        +{\bf a}({\bf p}-e{\bf A})\right) + e^2 {\bf a}^2\right).
    \label{H_ran_b}
\end{eqnarray}
The matrix elements $\langle Nk|H'|N'k' \rangle$ of $H'$ with the
eigenstates $|Nk\rangle$ of $H_0$ form a random matrix. Its
statistical properties can be compared to the those of the random
Landau matrix $\langle Nk|V|N'k' \rangle$ where $V({\bf r})$ is a
random electrostatic potential. The latter model has been extensively
studied and describes the transitions between integer QH plateaus
\cite{Huc95r}. We will show that the matrix elements of $H'$ have
similar statistics as those of $V({\bf r})$ in the limit of strong
average magnetic field $B_0$, despite the rather different nature of
the operators $H'$ and $V({\bf r})$. More precisely, we consider the
limit in which the fluctuations of the random magnetic field $b({\bf
  r}){\bf e}_z = \nabla\times {\bf a}({\bf r})$ are small compared to
the average field $B_0$ (we choose the average of $b({\bf r})$ to
vanish). In this limit the term quadratic in ${\bf a}$ can be
neglected in $H'$ and the coupling between different Landau levels of
$H_0$ becomes negligible. The intra-Landau-level matrix elements of
$H'$ are then given by
\begin{eqnarray}
  \langle Nk | H'| Nk'\rangle = \frac{\hbar e}{m^*} \Bigg(&&
  \frac{1}{2}\langle Nk |b({\bf r})| Nk'\rangle \nonumber \\
  && \mbox{} +
    \sum_{n=0}^{N-1} \langle nk |b({\bf r})| nk'\rangle \Bigg).
  \label{result}
\end{eqnarray}
This is our main result. It contains only matrix elements of the
gauge-invariant local magnetic field $b({\bf r})$. The quantity $\hbar
\omega_c({\bf r}) = \hbar e b({\bf r})/m^*$ is the deviation of the
local cyclotron energy from the average value $\hbar\Omega=\hbar
eB_0/m^*$. It follows from eq.~(\ref{result}) that in the lowest
Landau level $N=0$ the random magnetic field is indistinguishable from
a random electrostatic potential $V({\bf r})=\hbar \omega_c({\bf
  r})/2$, irrespective of the statistical properties of the random
magnetic field $b({\bf r})$, as has already been noted
\cite{KWAZ93}. When the magnetic field varies
sufficiently slowly on the scale of the cyclotron orbit radius
$R_c=(2N+1)^{1/2}l_0$, $l_0^2=\hbar/(eB_0)$, its matrix elements
become independent of the Landau level index and the random magnetic
field is strictly equivalent to the random electrostatic potential
$V({\bf r})=(N+1/2)\hbar \omega_c({\bf r})$. In both limits the random
magnetic field manifests itself only in the fluctuating cyclotron
energy in the $N$-th Landau level. We can thus apply all the known
results for electrostatic disorder to the present system. In
particular, the critical behavior is identical and the localization
length diverges in the center of the Landau levels with the same
exponent as in the IQHE. It further readily follows that the density
of states for a white noise distribution of the magnetic field in the
lowest Landau level is given by Wegner's result \cite{Weg83}. This is
in contrast to the situation in high Landau levels where the density
of states differs considerably from the electrostatic disorder case
\cite{AAMW95}. As the charge $e$ entering the Hamiltonian
(\ref{Hamiltonian}) is the charge of the electrons the critical
conductivity of the CFs is the same as that of the electrons as has
recently been seen experimentally \cite{Sea95}.

In ref.~\cite{KWAZ93} it has been argued that, in general, the random
magnetic field is equivalent to the potential $V({\bf r})=(N+1/2)\hbar
\omega_c({\bf r})$ plus gradient corrections. We can discuss this
statement if we express eq.~(\ref{result}) in terms of the Fourier
coefficients of $\omega_c({\bf r})$, $\omega_c({\bf
  r})=\sum_{\bf G} \omega_{\bf G} \exp (i{\bf G}\cdot {\bf r})$,
\begin{eqnarray}
  \label{fourier_h}
  \langle Nm | H'| Nm'\rangle &&= \hbar \sum_{\bf G} \omega_{\bf G}
  e^{-G^2l_0^2/2}
  G_{m,m'}\left(\frac{Gl_0}{\sqrt{2}}\right)\nonumber\\
  \times && \left[\frac{1}{2} L_N\left(\frac{G^2l_0^2}{2}\right) +
    \sum_{n=0}^{N-1} L_n\left(\frac{G^2l_0^2}{2}\right) \right],
\end{eqnarray}
where $m,m'$ are angular momentum quantum numbers in the symmetric
gauge ${\bf A}({\bf r})=B_0(-y{\bf e}_x+x{\bf e}_y)/2$ and
$G_{m,m'}(x)=(m'!/m!)^{1/2} x^{m-m'} L_{m'}^{m-m'}(x^2)$. An effective
electrostatic potential $V({\bf r})$, $V({\bf r})=\sum_{\bf G} V_{\bf
  G} \exp (i{\bf G}\cdot {\bf r})$, has the same matrix elements, if
\begin{eqnarray}
  \label{equiv_eff}
  V_{\bf G} L_N\left(\frac{G^2l_0^2}{2}\right) = \hbar \omega_{\bf G}
  \Bigg[ && \frac{1}{2}
    L_N\left(\frac{G^2l_0^2}{2}\right)\nonumber\\
  && \mbox{} + \sum_{n=0}^{N-1}
    L_n\left(\frac{G^2l_0^2}{2}\right) \Bigg].
\end{eqnarray}
We see that the effective electrostatic potential only exists if
$\omega_{\bf G}=0$ for $G^2l_0^2/2$ equal to the zeroes of $L_N(x)$.
For arbitrary random magnetic fields this is only fulfilled in the
lowest Landau level. In particular, there is no effective potential
for a white-noise distribution of the magnetic field in higher Landau
levels. If the magnetic field is sufficiently smooth, such that
$\omega_{\bf G}=0$ for $G^2l_0^2/2\geq x_N^{(1)}$, where $x_N^{(1)}$
is the first zero of $L_N(x)$, then the inverse of $L_N(G^2l_0^2/2)$
can be expanded into a power series in $G^2l_0^2/2$ and the effective
potential exists and can be written as a power series in
$-l_0^2\nabla^2$ acting on $\hbar \omega_c({\bf r})$, as claimed in
ref.~\cite{KWAZ93}.

Since, in general, the random magnetic field is not equivalent to an
electrostatic potential even in the limit of strong magnetic
field it is not evident that it has the same critical
behavior. According to eq.~(\ref{result})  the random matrix $\langle
Nk|H'|Nk' \rangle$ is equivalent not to a single random Landau matrix
but to a superposition of $N$ random Landau matrices, all containing
the same electrostatic potential but different Landau levels. This
leads to differences in physical properties like the density of
states. By studying the statistical properties of the matrix elements
of $H'$ and comparing them to those of an electrostatic random
potential we can argue that these differences are irrelevant for the
critical behavior of the system. To this end, we briefly review the
construction of the random Landau matrix for electrostatic
potentials. For gaussian correlations of a scalar potential $V({\bf
  r})$,
\begin{equation}
  \overline{V({\bf r})V({\bf r}')} =
  \frac{V_0^2}{2\pi\sigma^2} \exp
  \left(-\frac{|{\bf r}-{\bf r}'|^2}{2\sigma^2}\right),
\end{equation}
matrix elements $\langle Nk|V|Nk' \rangle$ in Landau gauge are
given in terms of uncorrelated random numbers $u(x,k)$,
$\overline{u(x,k)u(x',k')} = \delta(x-x') \delta_{k,-k'}$, by\cite{Huc95r}
\begin{eqnarray}
  \lefteqn{\langle N k_1|V|N k_2 \rangle =
    \frac{V_0\beta l_0}{\sqrt{2 L_y}\pi\sigma}
    \exp\left(-\frac{\kappa^2l_0^2\beta^2}{4}\right)}\nonumber\\
    & & \times\int d\xi\, u\left(\beta\xi +
    Kl_0,\kappa l_0\right)
  e^{-\xi^2}F^V_N(\xi,\kappa l_0;\sigma),
  \label{v_ran_mat}
\end{eqnarray}
where $K=(k_1+k_2)/2$, $\kappa=k_1-k_2$,
\begin{eqnarray}
  \lefteqn{F^V_N(\xi,x;\sigma) = \left(2^N N!\right)^{-1} \int d\eta
  \exp\left(-\frac{l_0^2+\sigma^2}{\sigma^2}\eta\right)}\nonumber\\
  & &\times H_{N}\left(\eta +
    \frac{\xi}{\beta}-\frac{x}{2}\right)
  H_{N}\left(\eta +
    \frac{\xi}{\beta}+\frac{x}{2}\right),
  \label{f_func}
\end{eqnarray}
$L_y$ is the width of the system, and $\beta^2=(\sigma^2 +
l_0^2)/l_0^2$. Using this result for gaussian correlations of the
magnetic field,
\begin{equation}
  \overline{b({\bf r})b({\bf r}')} =
  \frac{b_0^2l_0^2}{\sigma^2} \exp
  \left(-\frac{|{\bf r}-{\bf r}'|^2}{2\sigma^2}\right),
\end{equation}
the matrix elements of $H'$ are given by
\begin{eqnarray}
  \lefteqn{\langle N k_1|H'|N k_2 \rangle =
    \frac{V_0\beta l_0}{\sqrt{2 L_y}\pi\sigma}
    \exp\left(-\frac{\kappa^2l_0^2\beta^2}{4}\right)}\nonumber\\
    & &\times\int d\xi\, u\left(\beta\xi +
    Kl_0,\kappa l_0\right)
  e^{-\xi^2}F^B_N(\xi,\kappa l_0;\sigma),
  \label{b_ran_mat}
\end{eqnarray}
where $V_0^2 = 2\pi \hbar e b_0^2/\left( m^{*2} B_0\right)$, and
\begin{equation}
  F^B_N(\xi,x;\sigma) = \frac{1}{2}F^V_N(\xi,x;\sigma) +
  \sum_{n=0}^{N-1} F^V_n(\xi,x;\sigma).
  \label{g_func}
\end{equation}
Equations (\ref{v_ran_mat}) and (\ref{b_ran_mat}) differ only in the
weight functions $F^{V,B}_N$. In both cases these are polynomials of
degree $2N$ in $\xi$ and $x$. The critical behavior in the IQHE is
universal if it is the same for all polynomials $F^V_N(\xi,x;\sigma)$.
This has been numerically checked for the parameter combinations
$(N,\sigma)=(0,0)$, $(0,l_0)$, and $(1,l_0)$ \cite{Huc92}. For all
other values of $N$ and $\sigma$ universality in the IQHE is a
conjecture. The important feature of eqs.~(\ref{v_ran_mat}) and
(\ref{b_ran_mat}) seem to be the gaussian factors of $\xi$ and
$\kappa$ that reflect the Landau quantization, while the particular
form of the polynomial weight function seems to be rather irrelevant.
We therefore conjecture that random matrices of the form
(\ref{v_ran_mat}) and (\ref{b_ran_mat}) have the same critical
behavior for any weight function $F_N(\xi,x;\sigma)$ that is a
polynomial in $\xi$ and $\kappa$. This implies in particular that the
model under consideration belongs to the same universality class as
the IQHE.

We will now briefly derive the main result eq.~(\ref{result}). In a
complex notation for vectors in the $x$-$y$-plane, $z = x + i y$, we
can express the Hamiltonian
\begin{equation}
  H' = -\frac{e\sqrt{2}\hbar}{4m^*l_0}
  \left(\hat{a}_0\bar{a} + \hat{a}_0^{\dag} a \right) +
  \frac{e^2}{4m^*} a\bar{a} + \mathrm{h.c.},
\end{equation}
where $\bar{a}$ denotes the complex conjugate of $a$, in terms of
inter-Landau-level operators
\begin{displaymath}
  \hat{a}_0 = \frac{l_0}{\sqrt{2}\hbar}\left( \Pi_x^0 + i
    \Pi_y^0\right)\mbox{ and }
  \hat{a}_0^{\dag} = \frac{l_0}{\sqrt{2}\hbar}\left( \Pi_x^0 - i
    \Pi_y^0\right),
\end{displaymath}
where ${\bf \Pi}^0={\bf p}-e{\bf A}$. Using a Coulomb gauge for the
Chern-Simons vector potential
${\bf a}$,$\nabla \cdot {\bf a} = 0$, we have the following relations
between the commutators,
\begin{eqnarray}
  \left[\hat{a}_0^{\dag},a\right] + \Big[\hat{a}_0,\bar{a}\Big] &=&
  0,\\
  \left[\hat{a}_0^{\dag},a\right] - \Big[\hat{a}_0,\bar{a}\Big] &=&
  \sqrt{2}l_0 b(z),
\end{eqnarray}
so that $H'= H_1 + H_2 + H_3$, with \cite{note}
\begin{eqnarray}
  H_1 &=& \frac{1}{2} \hbar \frac{eb}{m^*},\\
  H_2 &=& - \frac{e\hbar}{\sqrt{2}ml_0} \left( \hat{a}_0^{\dag} a+
    \bar{a} \hat{a}_0 \right),\\
  H_3 &=& \frac{e^2}{2m^*} a\bar{a}.
\end{eqnarray}
The matrix elements of $H_3$ are on the average smaller by a factor of
$b_0/B_0$ than the matrix elements of $H_1$ and $H_2$ and can be
neglected in the limit $b_0\ll B_0$. Applying the Landau level ladder
operators $\hat{a}_0$ we get a recursion relation for the matrix
elements of $H_2$ (for clarity we suppress the dependence on the
intra-Landau-level quantum numbers $k$)
\begin{eqnarray}
  \langle N| H_2 | N \rangle &=& - \frac{\hbar e}{\sqrt{2}m^*l_0}
  \langle N | \hat{a}_0^{\dag} a + \bar{a} \hat{a}_0 | N \rangle \nonumber\\
  &=& \frac{\hbar e}{m^*}\langle N-1|b|N-1 \rangle +
  \langle N-1|H_2|N-1 \rangle \nonumber\\
  &=& \frac{\hbar e}{m^*} \sum_{n=0}^{N-1}\langle n|b|n \rangle,
\end{eqnarray}
thus leading to eq.~(\ref{result}).

In conclusion, we have studied charged quantum particles in a random
magnetic field in the limit that the fluctuations are much weaker
than the average magnetic field. This model arises in an
approximate treatment of the fermionic Chern-Simons theory of the
FQHE. We have shown that the model studied is strictly equivalent to an
electrostatic disorder potential in the two limits of the lowest
Landau level and of slowly varying magnetic field. A comparison of the
statistical properties of this model with known results for
electrostatic disorder led us to conjecture that these two models
have the same critical behavior. This implies that static fluctuations
of the Chern-Simons vector potential do not change the critical
behavior of ``composite fermions'' and that transitions between Landau
levels of the CFs belong to the same universality class as the IQHE.

Valuable discussions with J. Hajdu, M. Janssen, and M. Zirnbauer are
gratefully acknowledged. This work was performed within the research
program of the Sonderforschungsbereich 341.


\begin{thebibliography}{10}
\bibitem{note}
A similar result was obtained in ref.~\cite{KWAZ93}.
\end{thebibliography}
\end{document}